\documentclass[a4paper,11pt]{article}
\usepackage{pos}
\usepackage{float}
\usepackage{enumitem}
\usepackage{graphicx}
\usepackage{verbatim}
\usepackage[LGRgreek]{mathastext}
\usepackage[T1]{fontenc}
\usepackage[font=small,labelfont=bf,tableposition=top]{caption}
\usepackage[
backend=biber,
sorting=none,
style=numeric,
]{biblatex}
\title{A single photo-electron calibration system for the NectarCAM
camera of the Cherenkov Telescope Array Medium-Sized Telescopes}
 \ShortTitle{Calibration Technique for MST}

\author*[a]{Pooja Sharma}
\author[a,b]{Barbara Biasuzzi}
\author[a]{Jonathan Biteau}
\author[a]{Martin Bourgaux}
\author[c]{Sami Caroff}
\author[a]{Giulia Hull}
\author[a]{Michaël Josselin}
\author[a]{Kevin Pressard}
\author[b]{Patrick Sizun}
\author[a]{Tiina Suomijärvi}
\author[a]{Thi Nguyen Trung}

\affiliation[a]{Laboratoire de Physique des 2 Infinis, Irène Joliot-Curie, Université Paris-Saclay, Université de Paris, IN2P3/CNRS\\
  91405 Orsay, France}

\affiliation[b]{IRFU, CEA, Université Paris-Saclay\\
91191 Gif-sur-Yvette, France}

\affiliation[c]{LAPP, Univ. Grenoble Alpes, Univ. Savoie Mont Blanc, CNRS/IN2P3\\
74940 Annecy, France}

\forColl{CTA NectarCAM} % W/O "Collaboration"

\emailAdd{sharma@ijclab.in2p3.fr}

\abstract{This contribution aims to introduce the single photo-electron system designed to calibrate the camera of the Medium-Sized Telescopes of the Cherenkov Telescope Array (CTA). This system will allow us to measure accurately the gain of the camera's photodetection chain and to constrain the systematic uncertainties on the energy reconstruction of gamma rays detected by CTA. The system consists of a white painted screen, a fishtail light guide, a flasher and an XY motorization to allow movement. The flashes guided by the fishtail mimic the Cherenkov radiation and illuminate the focal plane under the screen homogeneously. Then, through the XY motorisation, the screen is moved across the entire focal plane of the NectarCAM camera, which consists of 1855 photo-multiplier tubes. In this contribution, we present the calibration system and the study on its optimum scan positions required to cover the full camera effectively. Finally, we illustrate the results of the calibration data analysis and discuss the performance of the system.} 

\FullConference{37$^{\rm{th}}$ International Cosmic Ray Conference (ICRC 2021)\\
		July 12th -- 23rd, 2021\\
		Online -- Berlin, Germany}

\DeclareCaptionLabelFormat{andtable}{#1~#2  \&  \tablename~\thetable}
\DeclareUnicodeCharacter{2212}{-}

\begin{document}
\maketitle

\section{Introduction}
The Cherenkov Telescope Array (CTA, https://www.cta-observatory.org/) is designed to detect indirectly gamma rays entering the Earth’s atmosphere by measuring the light induced by the development of atmospheric showers. Very-high-energy (VHE) gamma rays generate an electromagnetic shower of secondary electrons, positrons and photons. When these particles travel with a velocity that exceeds the velocity of light in the medium, Cherenkov radiation is produced. This radiation is reflected using large mirrors, towards a camera where it is captured by an array of Photo-multiplier tubes (PMTs). Owing to the thin and short flashes of Cherenkov radiation, cameras are designed to use few nanosecond exposure time to capture the sensitive signal. PMTs increase the intensity and convert the light into electrical signals which amplified, digitized and transmitted for further analysis. To detect this radiation, CTA is equipped with large, medium and small sized telescopes which together allow for significant improvement in sensitivity with respect to current generation instruments and an increased energy range from below 100 GeV to above 100 TeV. There will be tens of telescopes placed on complementary northern and southern sites (Canary Islands and Chile respectively) which promise access to any point in the night sky. Various science cases fall under the purview of CTA, as detailed in \cite{1}.

This paper focuses on the calibration system of the NectarCAM Camera \cite{3} for the Medium-Sized Telescopes (MST) which are designed to be sensitive in the range from at least 150 GeV to 5 TeV. The focal-plane calibration system will allow us to measure accurately the gain of the camera's photodetection chain and to constrain the systematic uncertainties on the energy reconstruction of gamma rays detected by CTA \cite{2}.
\vspace{-0.1cm}
\section{SPE Calibration system for NectarCAM}
\subsection{Overview}
Study of the incoming gamma rays involves reconstructing the gamma ray properties, namely the arrival time, direction and energy. To detect a low intensity light, we need to follow the conversion of photons to photo-electrons (p.e.), which are then multiplied to give the detected electrical signal. The gain is the factor by which the p.e. have been multiplied and it is required to measure the actual intensity of the light.

A dedicated single photo-electron (SPE) calibration system has been developed \cite{2} to measure the gain of each pixel of NectarCAM, one of the proposed cameras for MST. NectarCAM comprises 265 modules, each containing a set of 7 PMTs from Hamamatsu (R12992-100), with 7 dynodes each \cite{8}. In total, the camera comprises 1855 PMTs, in front of which a Winston cone is attached to ensure maximum light collection at the photocathode \cite{3}.

\subsection{SPE Calibration system}
In order to align the mirrors and perform point spread function (PSF) studies, a system which can move along the focal plane is required. This movable system can also be used for gain measurements when the camera is closed. The focal-plane calibration system mainly consists of:

\begin{itemize}[noitemsep,nolistsep]
    \item a screen, which is a 10 mm thick block of Poly(methyl methacrylate) (PMMA) which can cover 51 PMTs at once. Attached to it is the fishtail light guide with a groove for connection made up of the same material.
    \item a light box, which contains 12 light emitting diodes (LEDs) at 390 nm imitating Cherenkov light.
\end{itemize}

This calibration system is mounted inside the camera at a distance of 15 mm from the Winston cone on a system of XY rails attached to the motors which will move the system over the entire focal plane. When idle, the screen rests at the parking position i.e. on the bottom right corner of the camera where it does not cast a shadow on any of the pixels.

The light box source is used to inject light into the screen. To facilitate the injection of light, a fishtail light guide made up of PMMA is used. The screen is painted with highly reflective paint which causes the light to be reflected several times inside the screen. It is painted to serve dual purposes, the side of the screen facing the mirror dish of the MST is painted with three homogeneous layers of paint for Point Spread Function (PSF) and alignment studies. The other side is divided into 3 regions, each with layers of paint of varying thickness to allow for efficient spreading of light, whose transmissivity is used to measure the gain of the photo-detection chain in SPE regime.

\subsection{Diffuse-reflective screen}
An ideal screen should ensure homogeneous spreading of light, so that each PMT receives equal amount of light. To achieve this 3 main factors have been tested:

\begin{itemize}[noitemsep,nolistsep]
    \item Paint specification: this involves the careful study of painting patterns to ensure high reflectivity and homogeneous spreading. 
    \item Screen geometry: the screen should be as large as possible to enable a fast scan and should not cast a shadow on the focal plane when in parking position.
    \item Coating: dip-coating method is used to paint the screen and the fishtail. To ensure homogeneity and the reproducibility of the screen performance, a careful study of the dip-coating procedure was performed to limit the thickness of the paint layers on the screen. 
\end{itemize}

\subsection{Screen Test Setup}

To test the homogeneity of the screen, a dark room has been set up in IJCLab. Inside it, a NectarCAM front-end board (FEB) is connected to one PMT. The entire system is moved over the screen using motors on a system of rails (model ZLW-1040-S) to perform light intensity measurements. The intensity of LEDs can be adjusted using labview software interface. Several screens were tested using a high voltage of 1200 V for the PMTs and with appropriate voltage for LEDs to avoid saturation of light in bright spots of the screen.
\vspace{-0.3cm}
\subsection{Results}
Various tests were performed to study the effect of varying geometry, painting patterns and dip coating method in order to find a screen with maximum homogeneity. The final configuration consists of Eljen paint with $18\%$ dilution. The chosen painting pattern has 3 layers of paint between 0 and 16 cm from the light injection, 2 layers between 16 and 26 cm and 1 layer between 26 and 39 cm, which is the farthest region from injection point (configuration 3-2-1). A dedicated mask has been made to cover different regions of the screen while immersing the screen on the paint tank. For the final geometry, a flat octagonal screen of area 1338 cm$^2$, with flat fishtail light guide and a bent groove ($30^{\circ}$) has been selected. The bending allows the light to enter at an angle that results in more internal reflections and increased homogeneity. The withdrawal speed of the dip coating machine determines the adhesion of the paint on the screen acting against the gravity-induced viscous drag \cite{4}. The final screen is painted with a withdrawal speed of 2 mm/s.

Light intensity measurements were performed in high intensity regime ($ > 100$ p.e. produced at photocathode) by measuring the photon counts and producing histograms of charge integrated over 30,000 signals over a fixed time window (60 ns), at various points on the screen.  Finally, we obtained the homogeneity map shown in figure 1, measured over the screen with 10 contours ranging from half down to a tenth of the maximum screen emissivity. We can see that light is satisfactorily spread across the screen within 10 contours. The cumulative area covered by each contour along with the intensity lying within a given percentage of the maximum intensity for each contour is reported in Table 1.
\vspace{-0.2 cm}  
\begin{figure}[!ht]
  \centering
  \includegraphics[width=6.3cm, height=5.7cm]{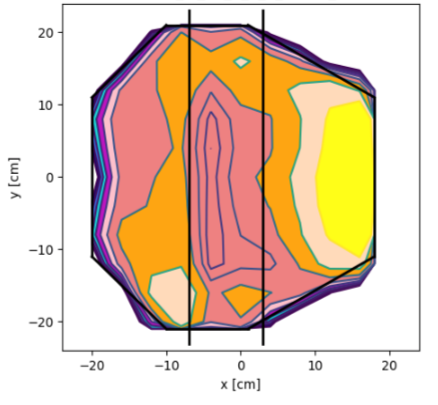}
    \hspace{0.2cm}
  \includegraphics[width=0.5cm, height=4.5cm]{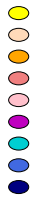}
  \qquad
  \scalebox{0.8}{
  \begin{tabular}[b]{cc}\hline
    Light intensity & Cumulative percentage \\
    (as a fraction of maximum) & of covered area \\ \hline
    $ > 50.0\%$ & 10 $\%$ \\ [3.1pt]
    $ 33.3\% > I > 50.0\%$ & 22 $\%$ \\ [3.1pt]
    $ 25.0\% > I > 33.3\%$  & 52  $\%$ \\ [3.1pt]
    $ 20.0\% > I > 25.0\%$ & > 90 $\%$ \\ [3.1pt]
    $ 16.7\% > I > 20.0\%$ & > 90 $\%$ \\ [3.1pt]
    $ 14.3\% > I > 16.7\%$ & > 90 $\%$ \\ [3.1pt]
    $ 12.5\% > I > 14.3\%$ & > 99 $\%$ \\ [3.1pt]
    $ 11.1\% > I > 12.5\%$ & > 99 $\%$ \\ [3.1pt]
    $ < 10.0\%$ & > 99 $\%$ \\ \hline
    \end{tabular}}
    \captionlistentry[table]{A table beside a figure}
    \captionsetup{labelformat=andtable}
    \caption{Figure 1 shows the light intensity contour map. The black lines mark the different regions of the screen which are painted with different layers. The different colours on the screen represent different contours which enclose the regions where the light intensity is within a given percentage of the maximum intensity measured on the screen, along with the cumulative area of screen enclosed by the contours is shown in Table 1.}
\end{figure}
\vspace{-0.8 cm}
\subsection{Calibration Scan}
With the screen ready, the optimal scan positions of the screen required to cover the entire focal plane need to be determined to ensure coverage of all pixels and minimise the time required for calibration. To achieve this, screen and camera geometry based on the Python library Sympy was used.
The screen is displaced across the camera horizontally with a fixed $step_{h}$ and vertically with a fixed $step_{v}$ ensuring that there is sufficient overlap between two successive screen coverage to account for edge effects. The limits of the movement are set by the edges of the camera internal frame. Once these limits have been set, we compute the time taken for each translation. To estimate this we have divided the total time into different parts. Figure 2 shows the representation of the various time periods considered for the movement of the screen, along with the equation of motions used for transition from one position to the next one. The transition from one position to the next one is given by accounting for:
\pagebreak
\begin{itemize}[noitemsep]
    \item $t_1$ - Time taken for the speed of the motor to ramp-up.
    \item $t_2$ - Time taken for the screen to move from one position to another.
    \item $t_3$ - Time taken for the speed of the motor to ramp-down.
\end{itemize}

In the equation provided in Figure 2, $a_1$, $a_2$ represent the ramp-up and ramp-down acceleration respectively. In addition we fix $step_h$, $step_v$ as 0.261 m and 0.313 m respectively.

\begin{minipage}{0.5\textwidth}
\begin{figure}[H]
\includegraphics[width=6.3cm, height=5cm]{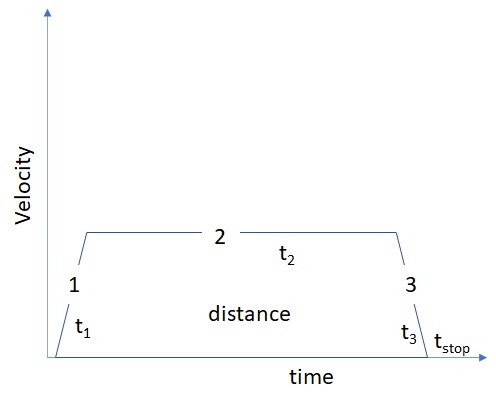}
\caption{Transit time periods.}
\end{figure}
\end{minipage} \hfill
\begin{minipage}{0.6\textwidth}
\begin{itemize}
\item $t_1 = \lvert\frac{v}{a_1}\lvert$
\item $ t_2 = (step_{h/v} - \lvert(\frac{1}{2} \times a_{1} \times t_{1}^{2})\lvert - \lvert(\frac{1}{2} \times a_{2} \times t_{3}^{2})\lvert)/v$
\item $t_3 = \lvert\frac{v}{a_2}\lvert$
\end{itemize}
\end{minipage}

\vspace{0.5 cm}
The total time estimate also accounts for the time taken for the screen to move from the parking position to the first position on the camera, transit from one column to an other and to return back to the parking position once the scan is over. We assume that 10,000 SPE flashes are shone upon the PMTs at a 1 KHz rate for each position, resulting in a rest time ($t_{stop}$) of 10 s. The minimum and the maximum velocity and acceleration of motors are set based on mechanical tests in the lab:

\begin{itemize}[noitemsep]
    \item v - constant velocity: [20, 50] mm/s
    \item $a_1$ - ramp-up acceleration: [20, 50] $mm/s^2$
    \item $a_2$ - ramp- down deceleration: [20, 50] $mm/s^2$ 
\end{itemize}

Finally, having considered all these factors, the total time taken for a complete scan is estimated to range from 17 to 28 minutes, for a total of 55 successive positions.

By using the information about the coordinate of the centroid of each translated position of the screen we can represent the screen positions on the camera geometry in Figure 3. The edges of the screen are shown in red, the coordinate of the centroid of each position of the screen is marked as a plus sign and the path of the screen is illustrated by dotted line. The limits imposed by the camera internal frame are signified by the black lines.

\begin{figure}[h]
\centering
\includegraphics[width=11cm, height=9.5cm]{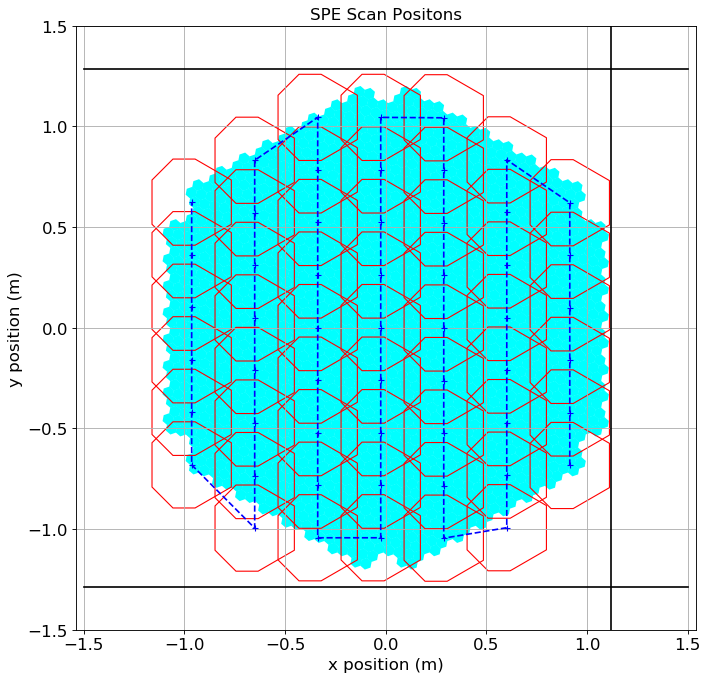}
\caption{Scan Positions required to calibrate the entire camera (see text).}
\end{figure}
\vspace{-0.5 cm}
\section{Test run with the calibration system}
\subsection{MST prototype \& calibration run in Adlershof}
In 2019, a mini NectarCAM camera consisting of 427 pixels was installed to the MST prototype telescopes in Adlershof (Germany). The main aim was to perform functional tests and tests of mechanical interfaces. Detailed study of the transport strategy and environmental constraints was also conducted. The very first data were acquired along with proper testing of the trigger and SPE system. The SPE screen was moved over all pixels with 4 LEDs powered at 10V. Data was acquired at 2kHz, for 20 seconds per position.

\subsection{Data analysis with Ctapipe}

The Ctapipe software was used to analyse the SPE data. Ctapipe\footnote{Karl Kosack et al.cta-observatory/ctapipe: v0.11.0. Version v0.11.0. May 2021.} is a low-level data processing framework for CTA which includes analysis of air shower images, processing of air shower images, such as reconstruction of incident air shower and discrimination between gamma-ray and hadronic showers \cite{6}. In Ctapipe, EventSource class iterates over all events and reads an event source which contains the various sub runs in the Aldershof dataset. Then EventSeeker is used to seek a particular event and access information about all events in one sub run. For each event a timestamp, identifying when a certain event occurred, is obtained. The time difference between consecutive timestamps is calculated. Finally, events with a time difference greater than 0.5 seconds are used to flag a change in the position of the screen during the scan. 
\vspace{-0.5cm}
\subsection{Results}

The charge of each event reconstructed after pedestal subtraction for every pixel is shown in figure 4. The reconstructed charge pattern on the camera corresponds to the emission pattern of the screen used to calibrate it (see \cite{3}). All events up to the event corresponding to the first time flag have been selected and charge for each event is extracted and displayed on the camera geometry.

\begin{figure}[H]
\centering
\centering
\begin{minipage}{0.5\textwidth}
  \centering
  \includegraphics[width=8cm, height=6cm]{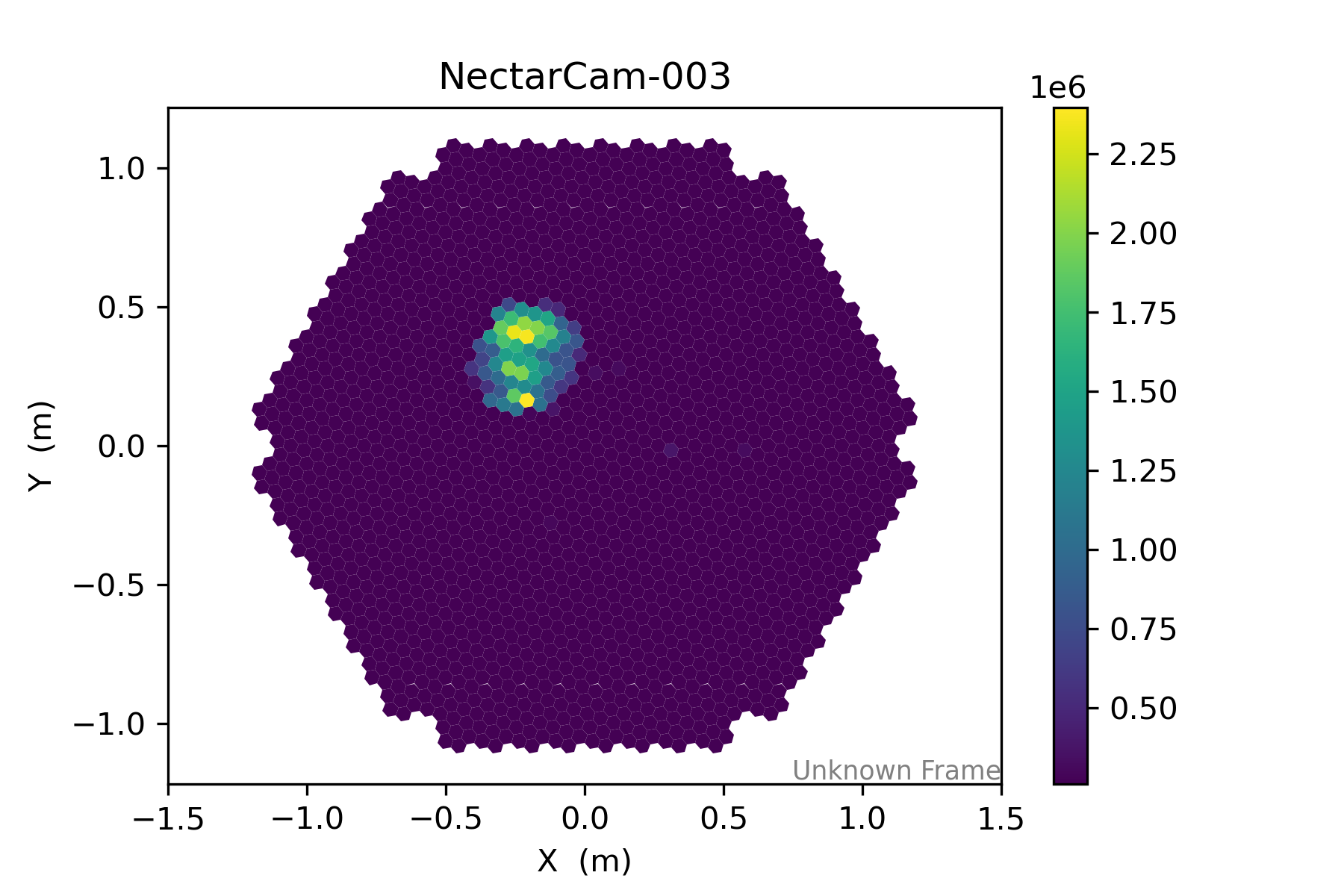}
  \captionof{figure}{Illuminated pixels in first position for \\ $\sim$ 20,000 events of data acquired in Aldershof}
  \label{fig:test1}
\end{minipage}%
\begin{minipage}{0.5\textwidth}
  \centering
  \includegraphics[width=8cm, height=6cm]{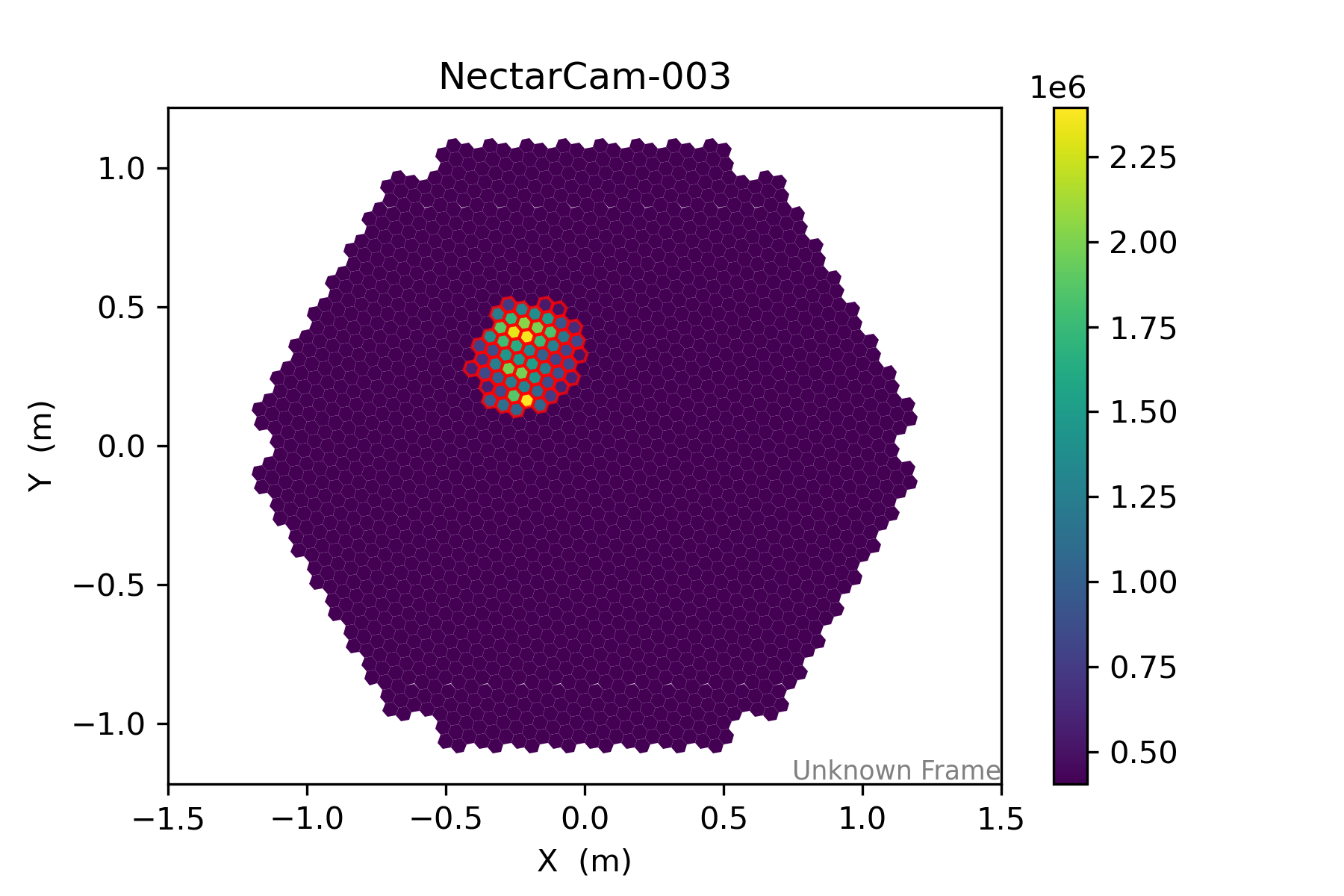}
  \captionof{figure}{Selection of illuminated pixels for $\sim$ 20,000 events of data acquired in Aldershof}
  \label{fig:test2}
\end{minipage}
\end{figure}

Events  which fall between consecutive transit periods correspond to a specific position of the screen during the scan. For a particular stop position of the screen, the waveforms of all events per pixel are summed to give a combined waveform. We use the average waveform in each pixel to determine the time when the signal is maximum ($T_{max}$). For the study we have used the integration time $T_{max} - 5 ns:T_{max} +15 ns$ to account for the part of the window where the signal is strong and noise is less. A similar window length is also used to subtract the pedestal from the waveform to further reduce the noise. The integrated waveform then gives the charge per event per pixel. Once the position of the screen is found, we select only the illuminated pixels as is shown in figure 5. For each illuminated pixel, a histogram of the charge distribution is stored.

The charge distribution of each pixel is modelled by a multi p.e. spectrum accounting for a two-gaussian SPE response \cite{7} and the pedestal level. Using this the gain of the PMTs can be computed. The first peak in Figure 6 and 7 corresponds to the pedestal (zero photon). Subsequent peaks correspond to photo-electrons, which can be reconstructed either in the high intensity region (Fig. 6) or in the low intensity region (Fig. 7) of the screen. Blue markers represent data points and orange curve is the fit. We obtain a mean gain of 56 ADC/p.e. and associated standard deviation of 2.1 ADC/p.e over the mini camera.

A histogram of the gain distribution of $\sim$ 21k events in all the illuminated pixels was also studied. This allowed us to constrain the pixel variation in gain to $\sim4\%$.

\begin{figure}[h]
\centering
\centering
\begin{minipage}{0.52\textwidth}
  \centering
  \includegraphics[width=8cm, height=6cm]{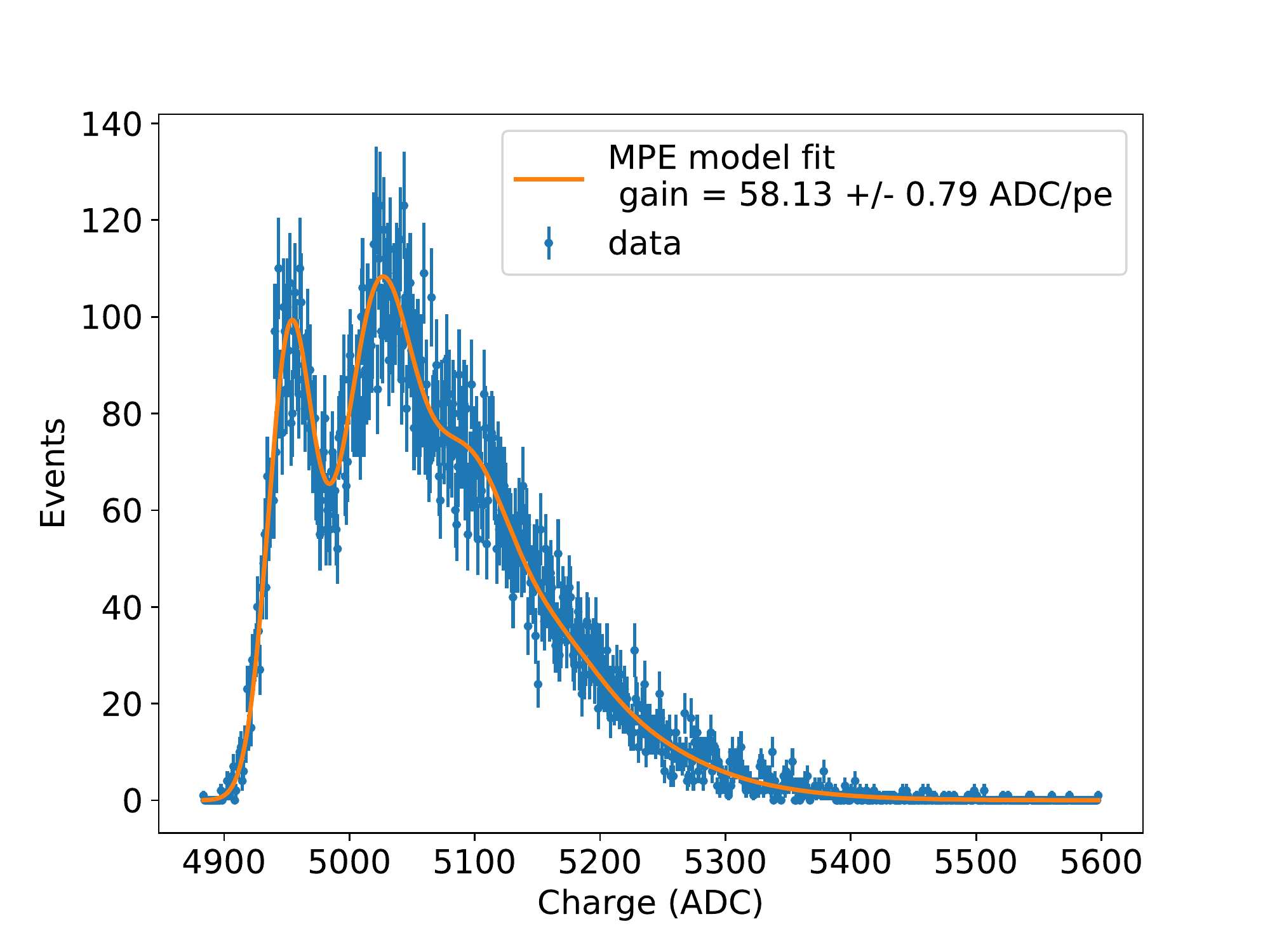}
  \captionof{figure}{SPE fit at high intensity region (Pixel 441)}
  \label{fig:test1}
\end{minipage}%
\begin{minipage}{0.52\textwidth}
  \centering
  \includegraphics[width=8cm, height=6cm]{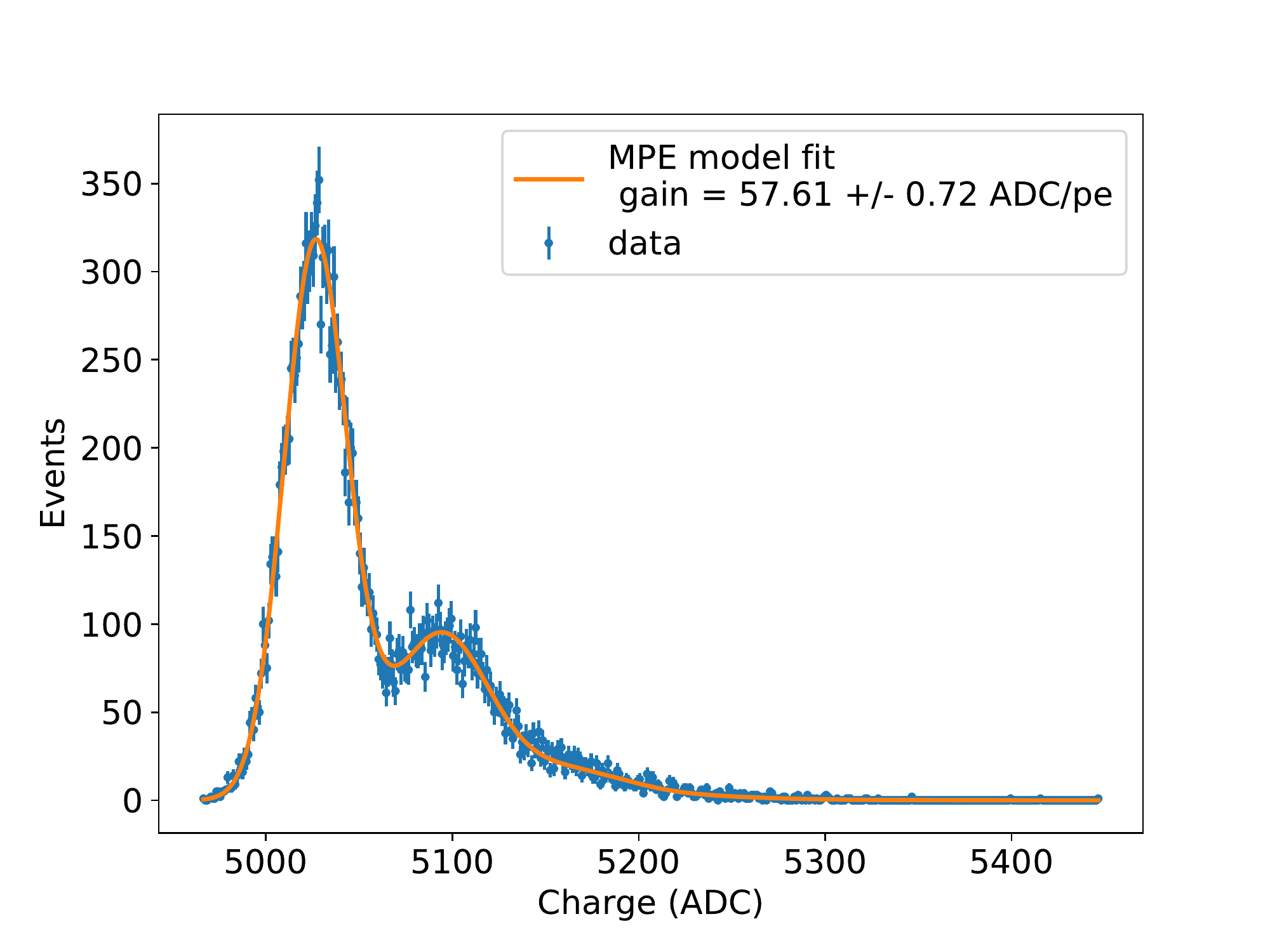}
  \captionof{figure}{SPE fit at low intensity region (Pixel 564)}
  \label{fig:test2}
\end{minipage}
\end{figure}

% \begin{figure}[H]
% \centering
% \includegraphics[width=12cm, height=7cm]{hist.png}
% \caption{SPE fit of real calibration run in Aldershof (see text)}
% \end{figure}
\vspace{-0.4cm}
\section{Conclusion}

In this paper, we have described the final design of the SPE calibration screen and its performance. In addition to this, the method to optimize the scan positions to reduce time taken to cover all PMTs has also been described. Data acquired with the SPE system integrated in NectarCAM camera that was installed on MST prototype was used for this study. By using this data, we have found a robust way to identify only those pixels which are illuminated by the screen to allow for a robust determination of the gain of the selected PMTs. The accuracy of gain determination is estimated to be $\sim4\%$. Data acquired with the fully equipped camera will enable to determine if an homogeneity of the gain better than $4\%$ can be achieved in the near future.

\section{Acknowledgement}
We would like to gratefully acknowledge the financial support from the agencies and organizations listed here: \url{http://www.cta-observatory.org/consortium_acknowledgments/}.
\vspace{-0.5cm}

\def\bibfont{\small}
\printbibliography[title={Bibliography}]

\end{document}